\documentclass[aps,prl,preprint,unsortedaddress]{revtex4-1}
\usepackage{graphicx}
\usepackage{amsmath}
\usepackage{epstopdf}

\begin{document}

%\preprint{}

\title{The observation of single Cooper-pair breaking by microwave light}

\author{N. J. Lambert}
\author{M. Edwards}
\author{A. A. Esmail}
\affiliation{Microelectronics Group, Cavendish Laboratory, University of Cambridge, Cambridge, CB3 0HE, UK}
\author{F. A. Pollock}
\affiliation{Atomic \& Laser Physics, Clarendon Laboratory, University of Oxford, Oxford, OX1 3PU, UK}
\author{S. D. Barrett}
\affiliation{Controlled Quantum Dynamics Theory, Imperial College London, South Kensington, London, SW7 2AZ, UK}
\author{B. W. Lovett}
\affiliation{SUPA, School of Physics and Astronomy, University of St Andrews, KY16 9SS, UK}
\affiliation{Department of Materials, Oxford University, Oxford, OX1 3PH, UK}
\author{A. J. Ferguson}
\affiliation{Microelectronics Group, Cavendish Laboratory, University of Cambridge, Cambridge, CB3 0HE, UK}
\email{ajf1006@cam.ac.uk}
\date{\today}

\begin{abstract}
We measure an aluminum superconducting double quantum dot and find that its electrical impedance, specifically its quantum capacitance, depends on whether or not it contains a single broken Cooper pair. In this way we are able to observe, in real time, the thermally activated breaking and recombination of Cooper pairs. Furthermore, we apply external microwave light and break single Cooper pairs by the absorption of single microwave photons.
\end{abstract}


\maketitle

A common way in which to detect light is to measure the electrical excitations created when light is absorbed in a material. In a semiconductor, an optical or infra red photon creates an exciton, which can be dissasociated, amplified and detected as an electrical current, for example, in an avalanche photodiode~\cite{Rieke2002}. At lower frequencies, it is convenient to measure the excitations created in superconductors, which have a smaller energy gap than typical semiconductors. In the commonly used kinetic inductance detectors, a far infra-red photon breaks multiple Cooper-pairs creating an ensemble of unpaired quasiparticle excitations which affect the electrical impedance of a superconducting resonator~\cite{Day2003}. The ultimate limit of such a detector would be to measure the pair of quasiparticle excitations that results from breaking a single Cooper pair. However, microscale superconducting resonators lack the necessary sensitivity. In this Letter we demonstrate the physical principle behind the detection of single microwave photons by the detected breaking of a single Cooper-pair in a superconducting nanostructure. 

A microwave frequency single photon detector would particularly find application in solid-state quantum information processing where it could, for example, be used to measure the occupancy of a superconducting resonator. Such a non-Gaussian measurement would enable a linear optical quantum computing~\cite{Knill2001} at microwave frequencies, where single photon sources can be generated using superconducting qubits~\cite{Leek2010}. Such a detector would also enable remote entanglement of stationary qubits through projective number state measurement of the photon field~\cite{Barrett2005}. Several approaches to single photon detection at microwave frequency have been tried~\cite{Shaw2009}, with the most promising based on the excitation of a higher lying state in a current biased Josephson junction~\cite{Chen2011}.

Superconducting Coulomb blockade devices are sensitive to single unpaired quasiparticles~\cite{Knowles2012,Maisi2013}. The best example of this is the pejoratively called 'quasiparticle poisoning', where the tunneling of an unwanted quasiparticle interrupts the desirable coherent behaviour of a superconducting qubit~\cite{Lutchyn2006,Catelani2011}. Quasiparticle poisoning can be monitored with a MHz bandwidth~\cite{Ferguson2006a} and recent innovations in microwave filtering have aimed to minimise the rate of such events~\cite{Corcoles2011,Barends2011,Saira2012}. An example closer to the present work is the use of quasiparticle tunneling to monitor the quasiparticle population in a superconductor that acts as the absorber for far-infrared light~\cite{Shaw2009,Stone2012}. This approach yields excellent performance that may improve upon kinetic inductance detection, but is not yet sensitive to a single created quasiparticle pair within the reservoir.

In order achieve sensitivity at the single pair level, our experiment employs an aluminium superconducting double dot (SDD). The SDD is made by triple angle thermal evaporation and controlled oxidation, and consists of two superconducting islands separated by a low resistance (7 k$\Omega$) tunnel barrier. Each island is also contacted to normal metal $\textrm{Al}_{0.98}$:$\textrm{Mn}_{0.02}$ leads by a high resistance (4 M$\Omega$) tunnel barrier (Fig.\ 1a).

For our low-temperature ($\sim0.1$ K) measurement, we embed the SDD in an LC circuit (Fig.\ 1b) and measure the complex reflection coefficient of a low-power (-121 dBm) radio-frequency signal at the circuit resonance ($f_0 = 349$ MHz) by homodyne detection at room temperature. This enables us to access the small changes in capacitance that occur due to charge transfer in the SDD. In particular, the device capacitance contains a fixed geometric contribution and a `quantum capacitance', given by $-\frac{d^2 E}{dV^2}$, where $E$ is the eigenstate energy and $V$ the potential on a capacitively coupled electrode~\cite{Duty2005,Sillanpaa2005}. 

\begin{figure}
\includegraphics{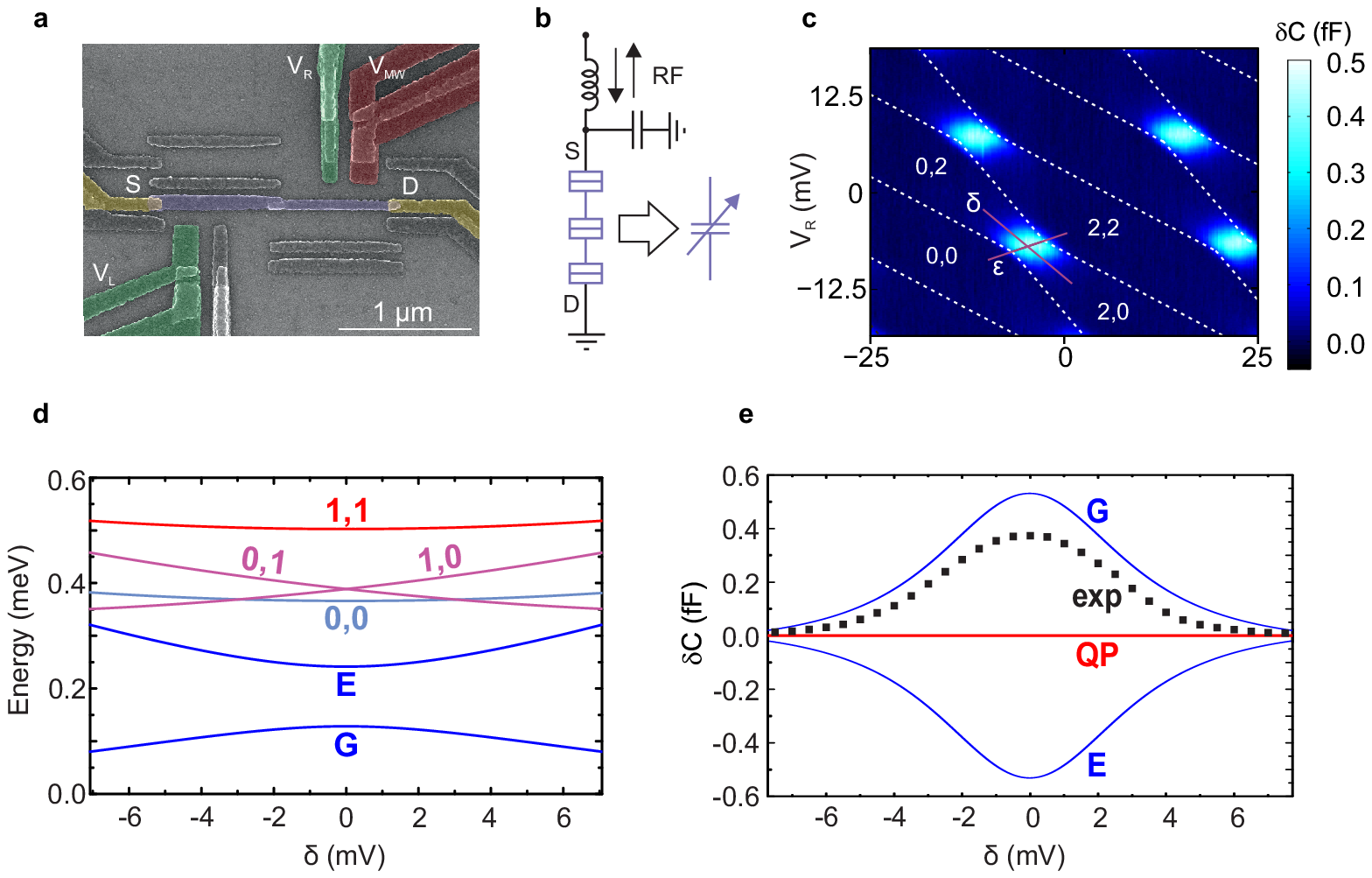}
\caption{\label{fig1}The superconducting double dot and quantum capacitance. (a) Scanning electron micrograph of a superconducting double dot device. The false colour regions show the dc gates (green), the microwave gate (red), the source and drain contacts (yellow) and the islands (purple). Uncoloured metal regions are artefacts of the triple angle evaporation process. (b) The device is embedded in an LC resonant circuit. The capacitance of the device depends on its charge state, which is apparent from the complex reflection coefficient of an incident signal at resonance. (c) Averaged measurement of capacitance as a function of the dc control gates. The charge states are shown and a reference state (0,0) is arbitrarily chosen. We illustrate the detuning axes $\delta$ and $\epsilon$. (d) Calculated bandstructure for $\delta$-detuning. Note that the (0,1) and (1,2) states (and the (1,0) and (2,1) states) are (approximately) degenerate and only one of each is shown.  (e) A comparison of measured value of $\delta C$ and the expected values for the ground and excited Cooper-pair bands. The bandstructure calculation was performed with experimentally determined values for the device energy parameters.}
\end{figure}

We alter the potentials of gate capacitors coupled to each of the superconducting islands and, from the circuit reflection coefficient, determine any capacitance shift ($\delta C$) due to the quantum capacitance. We plot $\delta C$ as a function of the gate potentials and average over multiple gate-potential ramps, defining $\delta C=0$ for the blockaded case e.g. in the middle of the (0,0) hexagon (Fig.\ 1c). 

In order to determine the capacitance signal generated by the different possible electronic states of the islands, in Fig.\ 1d we plot their energies as a function of a parameter $\delta$ which defines the difference in the two island potentials. The direction in the charging diagram of Fig.\ 1c along which $\delta$ lies is straightforwardly determined by looking at the shapes of the hexagons. The SDD has Cooper-pair ground states (e.g. (0,2) or (2,0)), which are split at their degeneracy point by Josephson tunnelling. The quantum capacitance is directly proportional to the second derivative of each state's energy with respect to the detuning, and so these two Cooper-pair states have equal and opposite $\delta C$ at degeneracy. In addition, there are excited states containing a different number of Cooper-pairs (e.g. (2,2), not shown), which can be reached through Andreev reflection from the Cooper pair ground states. These have constant curvature, and so $\delta C\approx0$.  There are also states with a quasiparticle pair (1,1) or single quasiparticles, e.g. (1,0). These cost an additional energy $\Delta$ per quasiparticle (at $B=0$, $\Delta=250$ $\mu$eV)~\cite{Lafarge1993,Tuominen1992}, but again have $\delta C\approx0$~\cite{Lambert2014b}.

From the charging diagram, we can also determine the energy scales of our device and we find the individual islands have charging energies $E_{C1}=314$ $\mu$eV and $E_{C2}=227$ $\mu$eV;  the interdot charging energy is $E_{CM}=88$ $\mu$eV~\cite{VanDerWiel2002}; and the interdot Josephson energy of $E_J=110$ $\mu$eV causes the splitting of the Cooper-pair bands. 

We proceed by applying an r.f.\ drive voltage between the source and drain, leading to an alternating contribution to $\delta$-detuning and a determination of $\delta C$. We compare this experimental $\delta C$ to a calculation based on the device parameters. We find good agreement (Fig.\ 1e): the small discrepancy between measured and calculated values can be accounted for by the finite occupation probability of the excited states. 

\begin{figure}
\includegraphics{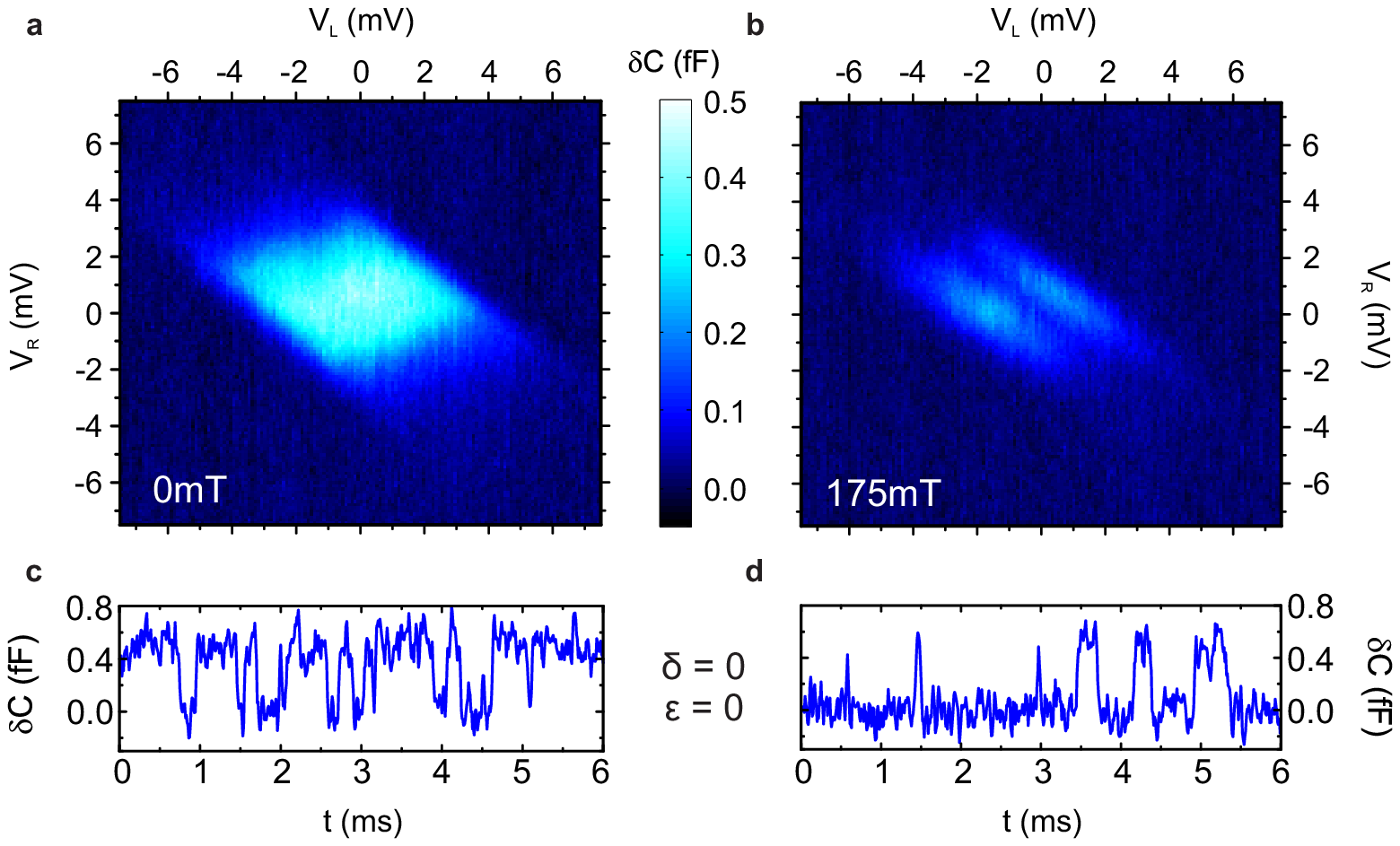}
\caption{\label{fig2}Excited state occupation. (a) \& (b) Zoom-in of capacitance showing the region around a single anti-crossing, for 0 mT and 175 mT.  (c) \& (d) Time domain measurements of capacitance for fixed gate voltages for 0 mT and 175 mT. These time domain traces are taken at the centre of (a) \& (b) i.e. at $\delta=0$, $\epsilon=0$.}
\end{figure}

We further characterize our device by applying a magnetic field. We find that the quantum capacitance signal at (0,2)/(2,0) degeneracy is reduced, indicating that a $\delta C=0$ state now becomes occupied (Fig.\ 2a \& b). To investigate this effect further we take, at fixed gate voltage, 10 s time-traces of the capacitance consisting of $10^6$ points. For both zero and finite magnetic field, the time-traces exhibit two-level switching between the ground Cooper-pair band  ($\delta C = 0.5 fF$) and an excited state with $\delta C=0$ (Fig.\ 2c \& d). At $B=0$ the ground Cooper-pair band is dominantly occupied, while at $B=175$ mT, the population is mostly in the $\delta C=0$ state. To determine $\Gamma_{G\rightarrow}$ and $\Gamma_{\rightarrow G}$, the rates for exiting and re-entering the ground Cooper-pair band, we divide the time trace into `up' and `down' intervals using a Schmitt trigger type algorithm. The two sets of periods are histogrammed, and exponential decays fitted to the histograms.

To determine which $\delta C=0$ processes contribute, we measure $\Gamma_{G\rightarrow}$ as a function of $\epsilon$-detuning (Fig.\ 3a), which corresponds to altering the overall energy of the SDD while maintaining a zero detuning of the two islands. At zero magnetic field, $\Gamma_{G\rightarrow}$ is constant for small $\epsilon$ but exponentially rises once a threshold is reached. The theoretical rate for G$\rightarrow$(1,1) is $\epsilon$-independent while the rates for G$\rightarrow$(0,1) and G$\rightarrow$(0,0) (and complementary processes) rapidly increase with $\epsilon$ (Fig.\ 3c \& d). On the basis of the $\epsilon$-detuning independence, we are able conclude that the pair-splitting process G$\rightarrow$(1,1) dominates near $\epsilon=0$. By contrast, at larger values of detuning, the G$\rightarrow$(0,1) and G$\rightarrow$(0,0) processes  dominate. As a magnetic field is applied, these processes which involve the leads have a greater contribution. At $B=175$ mT and $\epsilon=0$, the value of $\Gamma_{G\rightarrow}$ starts to contain contributions from excitation to other $\delta C=0$ excited states. We are able to conclude that our experiment is sensitive to individual pair-splitting events.

\begin{figure}
\includegraphics{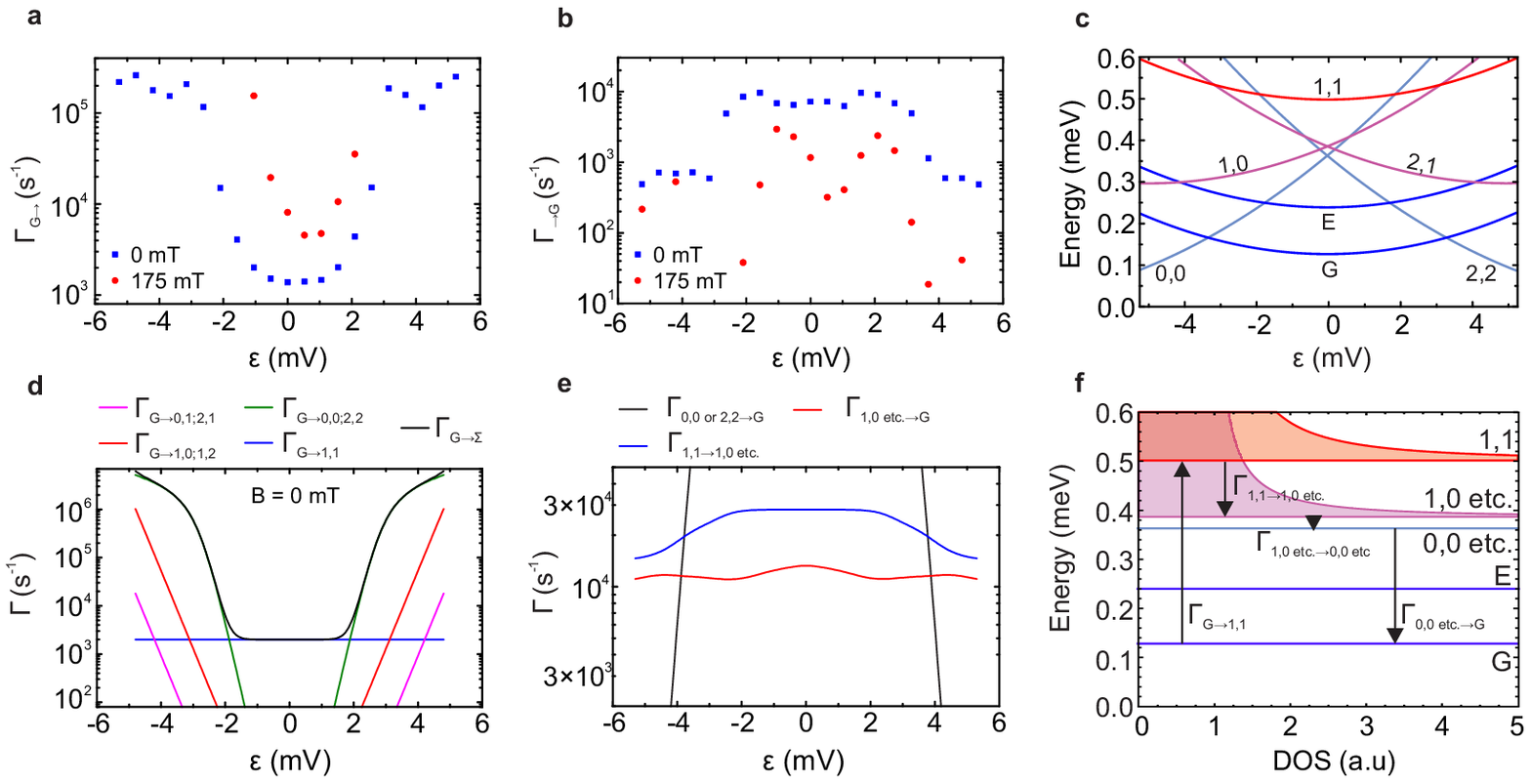}
\caption{\label{fig3}Pair-splitting and recombination dynamics. (a) \& (b) $\Gamma_{G\rightarrow}$ and $\Gamma_{\rightarrow G}$ measured as a function of the $\epsilon$-detuning for 0 mT and 175 mT. (c) Calculated bandstructure for $\epsilon$-detuning. (d) Theoretically predicted rates for $\Gamma_{G\rightarrow}$ including the rate for different processes and the total. These rates assume a temperature of 125 mK for the device and the environment-dependent rate $\Gamma_{G\rightarrow}$(1,1) is fitted to experimental data. (e) Theoretically predicted rates for the contributions to $\Gamma_{\rightarrow G}$. (f) Level structure at $\epsilon = \delta = 0$, showing the main generation/recombination cycle. The energy $E_{(1,1)-G}$ is 371 $\mu$eV, calculated from the measured values of $E_{C1,2,M}$, $E_J$ and $\Delta$.}
\end{figure}

We now consider the return rate to the ground state, measuring $\Gamma_{\rightarrow G}$, again as a function of $\epsilon$ (Fig.\ 3b). At small $\epsilon$ and $B=0$, $\Gamma_{\rightarrow G}$ is nearly constant. As $\epsilon$ increases, the (0,0) or (2,2) states, which also have $\delta C=0$, may become occupied, and these now lie below (0,2) and (2,0): therefore $\Gamma_{\rightarrow G}$ decreases. Occupation of these states can either occur after excitation to the (1,1) state or as part of, for example, a G$\rightarrow$(0,0)$\rightarrow$G cycle. For $B=175$ mT, there is a similar trend except that there is also a reduction in $\Gamma_{\rightarrow G}$ around $\epsilon=0$: it is this behaviour that leads to the (1,1) occupancy seen in the averaged measurement (Fig.\ 2b).

To explain the observed rates with the field applied, we consider first the recombination rate for a pair of quasiparticles in bulk aluminium corresponding to the SDD volume $(V\approx4\times10^{-3}$ $\mu\textrm{m}^3$), given by $\frac{8(1.76)^3}{\tau_0 V N_0 \Delta}\approx5$ kHz~\cite{Wilson2004,DeVisser2012}. This assumes an electron-phonon coupling constant $\tau_0=458$ ns and a single-spin density of states for normal state aluminium $N_0=1.72\times10^{10}$ $\mu\textrm{m}^{-3} \textrm{eV}^{-1}$. The equivalent direct recombination process, (1,1) $\rightarrow$ G, is dramatically suppressed from this bulk rate due to the forced spatial separation of the quasiparticles. Therefore we expect recombination to occur via particle exchange with the leads. Calculating the rates, we find that for small $\epsilon$ the (1,1) $\rightarrow$ (1,0) and similar processes occur rapidly (Fig.\ 3e). Subsequently, the (1,0) etc. $\rightarrow$ G decay, which can occur directly or via the (0,0) or (2,2) state, takes the SDD back to the ground state: The most common generation/recombination cycle for B=0 and zero $\epsilon$ are shown in Fig.\ 3f. The effect of magnetic field on the recombination cycle explains the reduction in $\Gamma_{\rightarrow G}$ at $B=175$ mT. At $B=175$ mT and $\epsilon=0$, the indirect (1,0) $\rightarrow$ (0,0)/(2,2) $\rightarrow$ G process is suppressed since the (1,0) state now lies lower in energy than the (0,0)/(2,2) state, hence thermal activation is required for the (1,0) $\rightarrow$ (0,0) transition. 

We next introduce a heavily attenuated microwave signal onto a gate capacitor in order to intentionally split Cooper-pairs, driving a cycle as illustrated in Fig.\ 4a. As before, we measure the time-domain switching signal (Fig.\ 4b) but this time under microwave illumination. The rate $\Gamma_{G\rightarrow}$ increases linearly with applied microwave power ($P$) and we parameterise the microwave sensitivity of the SDD by $\frac{d\Gamma_{G\rightarrow}}{dP}$ (Fig.\ 4c). The microwave sensitivity depends on the magnetic field; at a low field it has a constant value but above a threshold it exponentially increases (Fig.\ 4d). The magnetic field suppresses $\Delta$, reducing the energy difference between the (1,1) and ground states ($E_{(1,1)-G}$) and beyond a threshold field the microwave photons have sufficient energy for pair-splitting. We expect the dependence of microwave sensitivity on field to take the form of the superconducting density of states, however (as in Fig.\ 3a) transitions to the leads start to become important for large B, obscuring the peak in the density of states. As a result we only observe an exponential increase, which is a characteristic of the Dynes density of states near $\Delta$~\cite{Dynes1978,Pekola2010}. Measuring microwave sensitivity as a function of B for different microwave frequencies we find that the threshold depends on the applied microwave frequency, resembling $E_{(1,1)-G}(B)$ (Fig.\ 4d). The threshold for microwave induced pair splitting underestimates $E_{(1,1)-G}(B)$ due to the broadened density of states.

\begin{figure}
\includegraphics{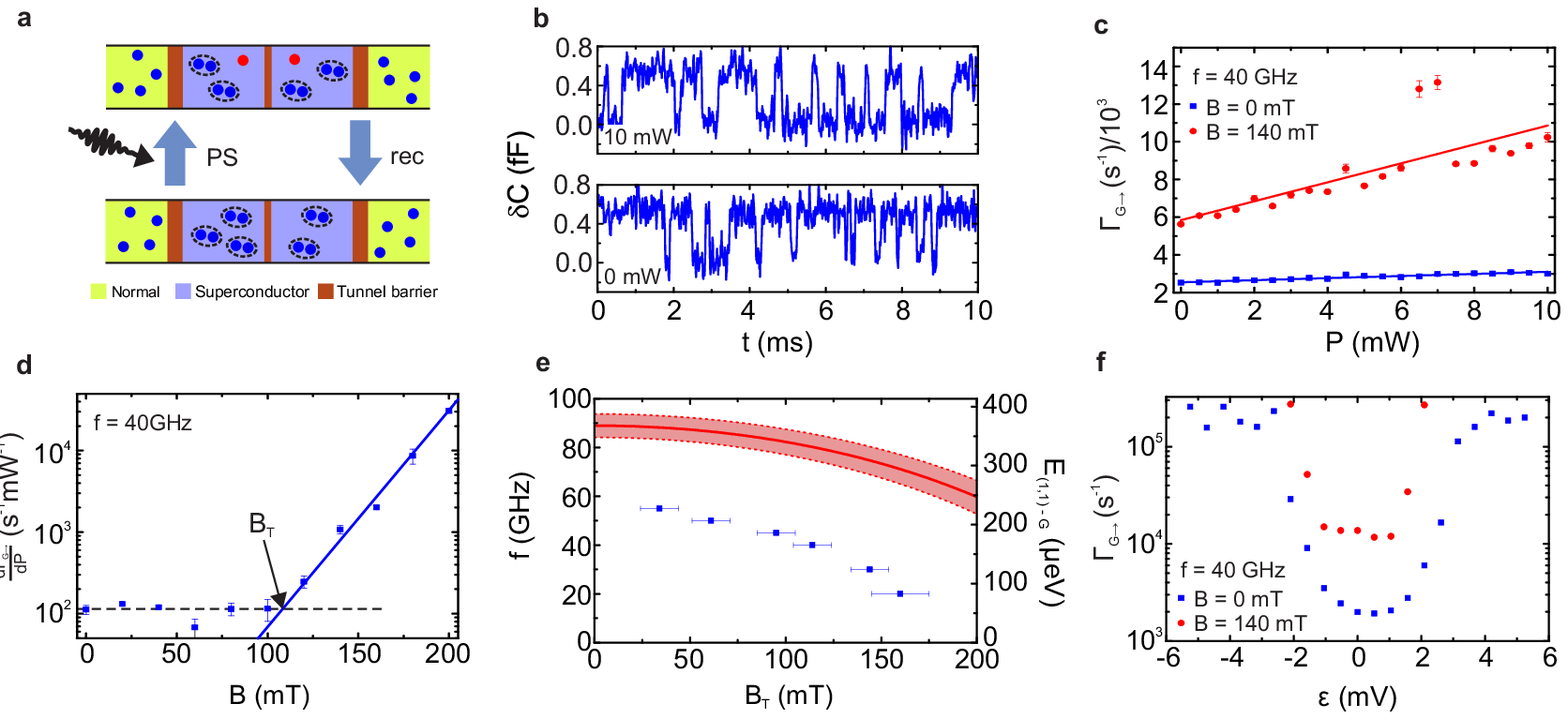}
\caption{\label{fig4}Microwave detection. (a) Absorption of a microwave photon or phonon induces pair-splitting (PS), taking the superconducting double dot from its even parity ground state to the excited (1,1) state, which has a pair of quasiparticle excitations. A recombination (rec) process returns the device to its ground state. (b) Time-domain traces measured under microwave illumination (at 40 GHz) for two different source powers. (c) $\Gamma_{G\rightarrow}$ as a function of microwave power at the signal generator, at fields of 0 mT and 140 mT, showing the change in gradient above and below threshold. (d) The microwave sensitivity, as a function of applied field, showing an exponential fit to the high field regime (solid line) rising above the background breaking rate (dotted line) at a threshold $B_T$. (e) Measured $B_T$ for different frequencies. Also plotted is $E_{(1,1)-G}$ as a function of magnetic field. (f) $\Gamma_{G\rightarrow}$  along $\epsilon$ with a 40 GHz microwave signal applied at 0 mT and 140 mT.}
\end{figure}

To confirm which ($\delta C=0$) state is excited by the microwave signal we measure sensitivity as a function of $\epsilon$, finding little $\epsilon$-dependence around $\epsilon=0$ and a subsequent increase for $\epsilon$-detuning (Fig.\ 4f). This indicates that around $\epsilon=0$ the (1,1) state is excited (as the G$\rightarrow$(1,1) transition is $\epsilon$-independent) while for larger detuning the $\epsilon$-dependent microwave assisted G$\rightarrow$(0,1) etc.\ and G$\rightarrow$(0,0) etc.\ transitions become important. This field dependence of microwave sensitivity with frequency, together with the linearity of $\Gamma_{G\rightarrow}$ with microwave power, indicates that a single photon process is responsible for pair-splitting.

Our paper has described single Cooper pair breaking and recombination at the single particle pair level and shown its sensitivity, at the single photon level, to microwave light. We suggest that future developments, such as impedance matching the double dot to a microwave feedline, could enable a useful detector based on this physical phenomenon.

\begin{acknowledgments}
A.J.F. would like to acknowledge the Hitachi Research fellowship, support from Hitachi Cambridge Laboratory and support from the EPSRC grant EP/H016872/1. B.W.L. is supported by a Royal Society University Research Fellowship. F.A.P. would like to thank the Leverhulme Trust for financial support.
\end{acknowledgments}

\bibliography{uWdetection}

\end{document}